\documentstyle[11pt]{article}
\begin{document}

\title{Finite matrix model of quantum Hall fluids on $S^2$}

\author{Yi-Xin Chen$^{a,b}$ \thanks{Email:yxchen@zimp.zju.edu.cn} 
, Mark D. Gould$^b$ and Yao-Zhong Zhang$^b$ \thanks{Email:yzz@maths.uq.edu.au}\\ 
$^a$Zhejiang Institute of Modern Physics, Zhejiang University,\\ 
             Hangzhou 310027, P. R.China,\\
$^b$Department of Mathematics, The University of Queensland, \\
    Brisbane Qld 4072, Australia}

\date{}
\maketitle

\begin{abstract}

\indent

Based on Haldane's spherical geometrical formalism of two-dimensional quantum Hall fluids, 
the relation between the noncommutative geometry of $S^2$ and the two-dimensional quantum Hall
fluids is exhibited. If the number of particles $N$ is infinitely large, two-dimensional 
quantum Hall physics can be precisely described in terms of the noncommutative $U(1)$ Chern-Simons theory 
proposed by Susskind, like in the case of plane. However, for the finite number of particles 
on two-sphere, the matrix-regularized version of noncommutative $U(1)$ Chern-Simons theory 
involves in spinor oscillators. We establish explicitly such a finite matrix model on two-sphere 
as an effective description of fractional quantum Hall fluids of finite extent. The complete sets 
of physical quantum states of this matrix model are determined, and the properties of quantum Hall 
fluids related to them are discussed. We also describe how the low-lying excitations are 
constructed in terms of quasiparticle and quasihole excitations in the matrix model. It is shown 
that there consistently exists a Haldane's hierarchical structure of two-dimensional quantum Hall 
fluid states in the matrix model. These hierarchical fluid states are generated by the parent fluid 
state for particles by condensing the quasiparticle and quasihole excitations level by level, 
without any requirement of modifications of the matrix model.

\vspace{.5cm}

\end{abstract}

\setcounter{equation}{0}

\section{Introduction}

\indent

The planar coordinates of quantum particles in the lowest Landau level of 
a constant magnetic field provide a well-known and natural realization of 
noncommutative space \cite{Dunne}. The physics of electrons in the lowest 
Landau level exhibits many fascinating properties. In particular, when the 
electron density lies in certain rational fractions of the density 
corresponding to a fully filled lowest Landau level, the electrons 
are condensed into special incompressible fluid states whose excitations 
exhibit unusual phenomena such as fractional charge and fractional 
statistics. For the filling fractions $\nu ={1 \over m}$, the physics of these 
states is accurately described by certain wave functions proposed by 
Laughlin \cite{Laughlin}, and more general wave functions may be used to 
describe the various types of excitations about the Laughlin states.

There has recently appeared an interesting connection between quantum Hall effect and noncommutative field theory. 
In particular, Susskind \cite{Susskind} proposed that noncommutative Chern-Simons theory on the 
plane may provide a description of the (fractional) quantum Hall fluid 
and, specifically, of the Laughlin states. Susskind's noncommutative 
Chern-Simons theory on the plane describes a spatially infinite quantum 
Hall system. It gives the Laughlin states at filling fractions $\nu$ 
for a system of an infinite number of electrons confined in the lowest 
Landau level. The fields of this theory are infinite matrices which act on 
an infinite Hilbert space, appropriate to account for an infinite number of 
electrons. Subsequently, Polychronakos \cite{Polychronakos} proposed a 
matrix regularized version of Susskind's noncommutative Chern-Simons 
theory in an effort to describe finite systems with a finite number of 
electrons in limited spatial extent. This matrix model was shown to 
reproduce the basic properties of the quantum Hall droplets and two special 
types of excitations of them. Furthermore, it was shown that there 
exists a complete minimal basis of exact wave functions for the matrix 
regularized version of noncommutative Chern-Simons theory at arbitrary 
level $\nu^{-1}$ and rank $N$, and that these are in one to one 
correspondence with Laughlin wave functions describing excitations of a 
quantum Hall droplet composed of $N$ electrons at filling fraction $\nu$ 
\cite{Hellerman}. It is believed that the matrix regularized version of 
noncommutative Chern-Simons theory is precisely equivalent to the theory 
of composite fermions in the lowest Landau level, and should provide an 
accurate description of fractional quantum Hall states. It does 
appear an interesting conclusion that they are in agreement with the long 
distance behavior, but the short distance behavior is different 
\cite{Karabali}. However, it should be pointed that the Polychronakos' finite matrix model is still defined 
on the two-dimensional plane.

It is well known that it is convenient to formulate the quantum Hall system on the two-dimensional sphere for 
the description of quantum Hall fluids. Such a formulation appears in the work on fractional quantum Hall 
effect based on the 
spherical geometry \cite{Haldane}. The Haldane's model is set up by a two-dimensional electron gas of $N$ particles on a 
spherical surface in radial monopole magnetic field. A Dirac's monopole is 
at the center of the two-dimensional sphere. The Haldane's model describes not only a variant of Laughlin's scheme 
with fully translationally invariant wave functions, but also a hierarchy of the quantum Hall fluid states.
It is a well known fact that the two-dimensional compact spherical space can be mapped to the flat Euclidean space 
by the standard stereographical mapping. In fixed limit, the connection between this model and the noncommutative 
Chern-Simons theory can be exhibited clearly. Precisely, the noncommutative property of particle's coordinates in the
Haldane's model should be described in terms of fuzzy two-sphere \cite{Madore} ( see below in details ). The different 
noncommutative manifolds should correspond to finite matrix models with different geometrical properties. Recently, there has 
been much interest in formulating Chern-Simons theories on noncommutative manifolds \cite{Morariu,Lizzi,Balachandran}. 
The present problem is what is the finite matrix model for the quantum Hall fluids on two-sphere.  

The goal of this paper is to establish the finite regularized-matrix model describing the quantum Hall fluids on $S^2$.
Based on the Hellerman and Raamsdonk \cite{Hellerman}'s discussion for the equivalence of two-dimensional quantum Hall 
physics and noncommutative field theory, one knows that the second-quantized field theoretical description of quantum 
Hall fluids for various filling fractions should involve in certain noncommutative field theories. On the 2-dimensional 
plane, such a
noncommutative field theory is the regularized matrix version of the $U(1)$ noncommutative Chern-Simons theory. For the
quantum Hall system on $S^2$, which is described by the Haldane's model, what we want here is to construct a finite matrix 
model for second-quantized field theory of quantum Hall fluids. On the other hand, we hope to explore the possible 
hierarchical structure of the quantum Hall fluids in the finite matrix model.

This paper is organized as follows. Section two introduces the 
two-dimensional quantum Hall model on the spherical geometry proposed by Haldane \cite{Haldane}, and analyzes 
the noncommutativity of the coordinates on $S^2$ by focusing on the lowest Landau level states of the system.
It will be shown that this noncommutative geometry is the geometry of fuzzy $S^2$. In order to establish the description of
effective field theory related with it, we introduce the Hopf mapping of this fuzzy geometry to perform the Hopf fibration.
Like the usual Hopf mapping of $S^2$ in the presnece of Dirac monopole field, the effective field theory is singular-free
on the field configurations obtained by the Hopf fibration with fuzzy $S^2$ as the base. These configurations are described 
by the spinor with two complex components. By taking the number of particles infinitely large, it is shown that this effective 
theory is equivalent to the $U(1)$ noncommutative Chern-Simons theory proposed by Susskind. However, our matrix regularized
version of the effective theory with the finite number of particles is different with the Polychronakos' finite matrix model,
and is related to the matrix fields of the spinor with two complex components since the effective theory is invariant 
under the $U(1)$ gauge thansformation of such spinors. 
In the section three, we provide the description of the finite matrix model of the quantum Hall fluids on $S^2$. 
Furthermore, the Fock space structure of this matrix model is analysed, and its complete sets of physical quantum states 
are determined. The properties of quantum Hall fluids related to them are also discussed. 
Section four investigates the condensate mechanism of the low-lying excitations in the finite matrix model of quantum Hall 
fluids on $S^2$. It is shown that there exists indeed the Haldane's hierarchy in the 2-dimensional quantum Hall fluids in our 
matrix model, and such hierarchy is dynamically generated by condensing of excitations of the quantum Hall fluids level 
by level. Section five includes a summary of the main results in this paper and remarks on further research in 
this direction.

\section{Haldane's quantum Hall system and fuzzy $S^2$ structure}

\indent

In the quantum Hall effect problem, it is advantageous to consider compact spherical space which 
can be mapped to the flat Euclidean space by the standard stereographical mapping\cite{Haldane}.
Haldane considered a system where a two-dimensional electron gas of $N$ particles is placed on a
two-sphere $S^2$ in a radial Dirac monopole magnetic field $B$. A point $x^a$ on $S^2$ with radius 
$R$ can be described by dimensionless vector coordinates $n^a =x^a/R$, with $a=1,2,3$ which satisfy 
$n^a n^a =1$. The single particle hamiltonian in this system reads 
\begin{equation}
{\cal {H}}= \frac{1}{2MR^{2}}\sum_{a}\Lambda^{a} \Lambda^{a}
\end{equation}
where $M$ is the effective mass, and ${\vec \Lambda}= {\vec r}\times [-i\hbar {\vec \bigtriangledown} -e{\vec A} ]
={\vec r}\times [{\vec p} -e{\vec A} ]$ is the dynamical angular momentum of the particle. The relation 
between the vector potential ${\vec A}$ and the magnetic field is given by ${\vec \bigtriangledown }
\times{\vec A}=B{\vec n}$. Due to the presence of the Dirac monopole field, the dynamical angular 
momentum $\Lambda^a$ does not obey the algebraic relation of the usual angular momentum. One can
easily check that they satisfy the commutation relations 
\begin{equation}
[\Lambda^a ,\Lambda^b ]=i\hbar \epsilon^{abc}(\Lambda^c +eBR^2 n^c ) .
\end{equation}
However, $L^a \equiv \Lambda^a -eBR^2 n^a$ provide the generators of rotations in the presence of 
the Dirac monopole field. Indeed, by direct calculation, one can show that
\begin{equation}
[L^a ,L^b ]=i\hbar \epsilon^{abc}L^c,~~~[L^a ,\Lambda^b ]=i\hbar \epsilon^{abc}\Lambda^c,~~~ 
[L^a ,n^b ]=i\hbar \epsilon^{abc}n^c      .
\end{equation}
The vector ${\vec \Lambda}$ has no component normal to the surface, so we have $L^a n^a =-eBR^2
=n^a L^a$. As pointed out by Haldane\cite{Haldane}, the spectrum of $\Lambda^a \Lambda^a $ determined by
the angular momentum operators $L^a$ is $\Lambda^a \Lambda^a =(L+eBR^2)^a (L+eBR^2)^a 
=\hbar^2 [l(l+1)-S^2 ]$. Because ${\vec \Lambda}$ is a hermitian operator and the hamiltonian
$H\sim \Lambda^a \Lambda^a $ must be larger than or equal to zero, one determines the spectrum of the
algebra $L^a$ as $l=S+n, n=0,1,2,\cdots$.
Hence, for a given $S$, the energy eigenvalues of the Hamiltonian Eq.(1) are 
\begin{equation}
E_{n} = \frac{\hbar^2}{2M R^2}[n(n+1) + (2n+1)S].
\end{equation}
The above energy spectrum when $n=0$ corresponds to the lowest Landau level. Since $S$ is the spin 
of the particle, the degenracy of the lowest Landau level is $2S+1$.

On the other hand, we can discuss the classically canonical dynamics from the hamiltonians $H$
and $H+V(x^a )$, where $V(x^a )$ is the potential energy with rotational symmetry. By 
means of the correspondence between classical and quantum physics, one can 
straightforwardly read off the fundamental Poisson brackets of the classical degrees of freedom from
their corresponding commutation relations. In the canonical hamiltonian formulation, the evoluton
of dynamical variables with time is described by the canonical Hamilton equation, i.e., ${\dot \Lambda^a}
=\{\Lambda^a, H\}=\frac{eB}{M}\epsilon^{abc}\Lambda^b n^c {\not =} 0$. This implies that the dynamical 
angular momentum is not a conservative quantity of the system. In fact, in the presence of the Dirac 
monopole field, the generator of rotations is modified to $L^a$, which is a conservative quantity since
${\dot L^a}=\{L^a ,H\}=\frac{-1}{MR^{2}}\epsilon^{abc}\Lambda^{b} \Lambda^{c}=0$. If we consider the system
including a term of potential energy $V(x^a)$ with the symmetry of rotations, $L^a$ is still conservative.
That is
\begin{equation}
{\dot L^a}={\dot \Lambda^a} -eBR^2 {\dot n}^a=0 .
\end{equation}
So the variation of $n^a$ with time can be given by the canonical hamiltonian equation of $\Lambda^a$
\begin{equation}
{\dot n}^a=\frac{1}{eBR^2}[\frac{eB}{M}\epsilon^{abc}n^b \Lambda^c 
+\frac{\partial V}{\partial n^b}\epsilon^{abc}n^c ].
\end{equation}
Since we are interested in the equation of motion in the lowest Landau level, we can take the infinite limit
of mass $M\rightarrow \infty $. In this limit, we obtain the following equation of motion
\begin{equation}
{\dot n}^a=\frac{1}{eBR^2} 
\frac{\partial V}{\partial n^b}\epsilon^{abc}n^c .
\end{equation}
This implies that the momentum variables can be fully eliminated in the lowest Landau level. The elimination 
of momentum variables leads to the coordinates on the two-sphere which are noncommutative. Restricted to the
lowest Landau level states, the equation of motion can be equivalently derived from the fundamental Poisson
bracket
\begin{equation}
\{n^a ,n^b \}=\frac{1}{eBR^2}\epsilon^{abc}n^c ,~~~~n^a n^a =1.
\end{equation}
This Poisson algebra can be realized by the matrix commutator
\begin{equation}
[n^a ,n^b ]=\frac{1}{eBR^2}i\epsilon^{abc}n^c ,~~~~n^a n^a =1.
\end{equation}
Conclusively, if we focus on the lowest Landau level of the system, the Haldane's spherical 
geometry becomes the noncommutative geometry of the fuzzy $S^2$\cite{Madore}.

In order to exhibit the fuzzy property of algebra (9), we finish the isomorphic mapping 
from algebra (9) to the $SU(2)$ algebra. Set $n^a \mapsto \frac{X^a}{eBR^2}$, so the 
equation (9) becomes
\begin{equation}
[X^a ,X^b ]=i\epsilon^{abc}X^c
\end{equation}
which is the standard $SU(2)$ algebra. The quadratic Casimir of $SU(2)$ in the $N$-dimensional 
irreducible representation is given by
\begin{equation}
X^a X^a =\frac{1}{4}(N^2 -1).
\end{equation}
The constraint $n^a n^a =1$ leads to
\begin{equation}
eBR^2 =\sqrt{\frac{1}{4}(N^2 -1)}.
\end{equation}
This relation has shown that the two parameters $B$ and $R$ should be quantized, which exhibits 
the fuzzy property of two sphere. In order to compare them with the usual expression, we rewrite
the relation $eBR^2$ as $eBR\cdot R\equiv \frac{R}{\theta^{\prime}}=\sqrt{\frac{1}{4}(N^2 -1)}$,
where $\theta^{\prime}=eBR$. Then, we have
\begin{equation}
[n^a ,n^b ]=i\frac{\theta^{\prime}}{R}\epsilon^{abc}n^c \equiv i2\theta\epsilon^{abc}n^c   ,~~n^a n^a =1.
\end{equation}
This algebraic relation is the starting point of the following discussion about the Hopf mapping of 
the fuzzy $S^2$.

It is well known that for a monopole field, no single vector potential exists which is singularity-free 
over the entire manifold $S^2$. The use of two vector potentials living respectively on the north and on 
the south semi-spheres, which was advocated by Wu and Yang\cite{Wu}, provides a way round the 
singularity problem, since one can use each in a region where it is singularity-free, and then connects 
the two in a convenient overlap region by a gauge transformation. However, the Wu-Yang procedure is not 
well adapted for our later purpose to establish the efffective description of the system and its 
quantization. For the case of $U(1)$ Dirac's monopole, one can obtain the related effective Lagrangian, 
which is singularity-free, by using the Balachandran formalism\cite{Balachandran1}. The key step is to 
finish the first Hopf fibration of 
$S^2$ to get $S^3$. The first Hopf map is a mapping from $S^3$ to $S^2$ and is related to Dirac's monopole. 
In the presence of a Dirac's monopole, the $U(1)$ bundle over $S^2$ is topologically non-trivial. However, since 
$S^3$ is parallelizable, one can use first Hopf map to define a non-singular vector potential due to Dirac's 
monopole everywhere on $S^3$, called as the first Hopf fibration. 

Let us introduce the notation $z=\left ( \begin{array} {l}
z_1 \\
z_2 \end{array} \right )$ for the two-component spinor. The spinor $z$, in principle, has three degrees of freedom
since the normalization condition $z^{\dagger}z=1=|z_1|^2 +|z_2|^2$ is the only constraint on the two complex
numbers $z_1 ,z_2$. So they are actually defined on the surface of $S^3$. However, the Hopf projection map 
which takes us from $S^3$ to $S^2$ is given by
\begin{equation}
{\vec n}=z^{\dagger}{\vec \sigma}z ,
\end{equation}
where $\sigma^a$ are three Pauli matrices. It should be noticed that the $U(1)$ transformation 
$z\rightarrow e^{i\alpha }z$ leaves $n^a$ invariant and so the inverse image of any point on $S^2$ is a circle 
on $S^3$. Now we ask what Poisson relation for $z$'s can be used to produce the Poisson algebra (8) of the fuzzy 
$S^2$. It can be easily checked that the answer to this problem is
\begin{equation}
\{z,z^{\dagger}\}=\theta ,
\end{equation}
that is,
\begin{equation}
\{z_1 ,{\bar z}_1\}=\theta =\{z_2 ,{\bar z}_2\}, \{z_1 ,{\bar z}_2\}=0=\{z_2 ,{\bar z}_1\}.
\end{equation}

Subsequently, we focus on the description of the effective action of particles in the presence of the
Dirac monopole field. In the presence of a Dirac monopole, the $U(1)$ bundle over $S^2$ is topologically 
non-trivial. Such a non-trivial topological character leads to the appearance of an additional term, called
the Wess-Zimino term, in the effective action of the system. The effective action had been obtaned 
by Stone\cite{Stone} in the discussion of the coupling of the $SO(3)$ rotor and spinor and the calculation of the 
Berry phase. The effective action reads
\begin{equation}
I=\frac{1}{2f}\int dt\dot{n^a}\dot{n^a} +\int {\cal A}^a dn^a ,
\end{equation}
where, $f$ is dependent of the parameters of the system, and ${\cal A}$ is the potential of the Dirac monopole which
cannot be globally expressed on $S^2$ due to the singularity of the Dirac string. However, by means of the 
Hopf fibration of $S^2$ and its $U(1)$ gauge symmetry, the potential of the Dirac monopole can be globally written on
$S^3$ as
\begin{equation}
{\cal A}=i\frac{\tilde \lambda}{2}[z^{\dagger}dz-dz^{\dagger}z],
\end{equation}
where ${\tilde \lambda}$ is related to the magnetic charge of the Dirac monopole.
It should be pointed out that the potential ${\cal A}$ is equivalent to ${\cal A}^a dn^a $ up to an $U(1)$ gauge 
transformation, and is non-singular everywhere on $S^3$. Furthermore, the first term in the effective action (17)
can be also described by the spinors defined on $S^3$\cite{Balachandran1,Aitchison}. In fact, if one quantizes the
sytem described by the effective action after finishing the Hopf fibration, he gets the energy spectrum determined
by the hamiltonian (1)\cite{Aitchison}. Hence, the Haldane's quantum Hall system on the spherical geometry can be
equivalently described by the effective action (17). Restricted on the lowest Landau level state, as mentioned by 
us above, the contribution of kinetic energy in the effective action should be ignored, which is equivalent to taking the
infinite limit of $f$. So the physics in the lowest Landau level is described by the following action
\begin{equation}
I_{e}=\int {\cal A}.
\end{equation}

As mentioned above , the $U(1)$ gauge transformation $z\rightarrow e^{i\alpha}z$ leaves $n^a$ invariant, so
the effective action after finishing the Hopf fibration is also invariant under such an $U(1)$ gauge transformation.
Furthermore, projected in the lowest Landau level state, the effective action $I_{e}=i\frac{\tilde \lambda}{2}
\int dt[z^{\dagger}\partial_t z-\partial_t z^{\dagger}z]$ should possess this $U(1)$ gauge symmetry. By the standard
way of introducing the coupling of gauge field, we can write the effective action in the explicitly gauge invariant
form $I_{e}=i\frac{\tilde \lambda}{2}\int dt[z^{\dagger}(\partial_t +iA_{0})z-(\partial_t -iA_{0}) z^{\dagger}z]$, where
$A_{0}$ is a $U(1)$ gauge field. Indeed, this action is invariant under the $U(1)$ gauge transformations 
$z\rightarrow e^{i\alpha}z$ and $A_0 \rightarrow A_0 -\partial_t \alpha $. However, now the spinor $z$ becomes 
noncommuattive since it is from the Hopf mapping of the fuzzy $S^2$. The matrix realization of $z$ is required by the
non-trivial algebraic relations (15) [i.e. (16)]. The gauge field $A_0$ should adjointly act on the matrix $z$ in order to 
make the covariant derivative $\partial_t +iA_0$ satisfy the derivative property. Finally, the effective action
projected in the lowest Landau level state is given by
\begin{equation}
I_{e}=i\frac{\tilde \lambda}{2}\int dt Tr\{z^{\dagger}(\partial_t z +[A_{0},z])-
(\partial_t z^{\dagger} -[A_{0}, z^{\dagger}]z\}.
\end{equation}
This is a matrix theory similar to that describing $D0$-branes in the string theory\cite{Banks,Bernevig}. We can
use this theory to investigate the fluctuations of the spherical brane, which describes the excitations of the Hall 
fluids living on $S^2$, by using the method of expanding the matrix field in terms of the fluctuations around the classical
configurations\cite{Bernevig}.

First of all, let us introduce $\xi^r , r=1,2$, as the parameterizing coordinates of the $S^2$. The transformations of 
area preserving diffeomorphisms on this two-dimensional space are given by
\begin{equation}
\xi^r \rightarrow \xi^r +\beta^r (\xi),~~~\partial_r (w(\xi)\beta^r (\xi))=0,
\end{equation}
where $\beta^r $ can be locally written as
\begin{equation}
\beta^r (\xi)=\frac{\epsilon^{rs}}{w(\xi)}\partial_s \beta(\xi),
\end{equation}
and $w(\xi)$ is a 2-dimensional measure for the normalization.
The transformation rules of the fields are determined by introducing the Poisson brackets defined with respect to the 
measure $w(\xi)$ as 
\begin{equation}
\{A,B\}=\frac{\epsilon^{rs}}{w(\xi)}\partial_r A\partial_s B.
\end{equation} 
The transformations of the fields are $\delta X^a =\{\beta , X^a\}$ and $\delta A=\partial_t \beta 
+\{\beta ,A\}$. The coordinates $\xi^r , r=1,2$, parameterize not only the fields $n^a$ on $S^2$ but also the spinor 
field $z$ through the Hopf mapping ${\vec n}=z^{\dagger}{\vec \sigma}z $. However, in order to describe the consistent 
dynamics of the system, the definition of Poisson bracket (23) should coincide with that of the fundamental Poisson
brackets (15). Comparing (23) with (15), we get 
$W(\xi)=\frac{2}{\theta}det|\frac{\partial(x^1 ,x^2 )}{\partial(\xi^1 ,\xi^2)}|
=\frac{2}{\theta}det|\frac{\partial(x^3 ,x^4 )}{\partial(\xi^1 ,\xi^2)}|\equiv \theta^{-1}W$, where $z_1 =x^1 +ix^2$ and 
$z_2=x^3+ix^4$. According to the transformation rule of the fields, we have
\begin{equation}
\delta x^i =\frac{\theta }{W}\epsilon^{rs}\partial_r \beta \partial_s x^i
=\frac{\theta }{W}\epsilon^{rs}\partial_j \beta\partial_r x^j \partial_s x^i
=\theta \epsilon^{ij}A_{j},
\end{equation}
and
\begin{equation}
\delta x^{\tilde i} =\frac{\theta }{W}\epsilon^{rs}\partial_r \beta \partial_s x^{\tilde i}
=\frac{\theta }{W}\epsilon^{rs}\partial_{\tilde j} \beta\partial_r x^{\tilde j} \partial_s x^{\tilde i}
=\theta \epsilon^{{\tilde i}{\tilde j}}A_{\tilde j},
\end{equation}
where $i,j=1,2$ and ${\tilde i},{\tilde j}=3,4$. The above transformation relations should be understood as the matrix 
variables expanded in terms of the fluctuations $A$  around the classical solutions $x^{i(0)}$ and $x^{{\tilde i}(0)}$, 
which determine the classical spinor solution $z^{(0)}$. These classical solutions $z^{(0)}$ and $z^{\dagger(0)}$ obey the 
fundamental Poisson relations (15). Substituting the matrix variable expanssions with fluctuations into the
effective action $I_{e}$, we get
\begin{eqnarray}
I_{e} ={\tilde \lambda }\int dt Tr\{2\theta A_0 & + & \theta^2 \epsilon^{\mu \nu \lambda }
(A_{\mu}\partial_{\nu}A_{\lambda}
+\frac{2}{3}A_{\mu}A_{\nu}A_{\lambda})\nonumber\\ 
& + &\theta^2 \epsilon^{{\tilde \mu}{\tilde \nu}{\tilde \lambda}}
(A_{\tilde \mu}\partial_{\tilde \nu}A_{\tilde \lambda}
+\frac{2}{3}A_{\tilde \mu}A_{\tilde \nu}A_{\tilde \lambda})\}.
\end{eqnarray}

The $x^{i(0)}$ and $x^{{\tilde i}(0)}$ are the matrices of the classical solution to be related with the 
noncommutative coordinates of the fuzzy $S^2$ by means of the Hopf mapping.
Since any matrix can be expressed in terms of finite sum of products 
$\prod_{i{\tilde i}}exp\{ip_ix^{i(0)}\}exp\{ip_{\tilde i}x^{b(0)}\}$,
 the $N\times N$ matrices $A_{\mu}$  and $A_{\tilde \mu}$  can be thought of as functions of  
$x^{i(0)}$ and $x^{{\tilde i}(0)}$. 
Based on this fact, we can pass the effective Lagrangian to the continuum limit by taking $N$ large. The changes of the 
coordinates $\xi^r, r=1,2$ parameterizing the spherical geometry induce the variations of the matrix fields 
$x^{i(0)}$ and $x^{{\tilde i}(0)}$. In the continuum limit, the $N\times N$ matrices $A_{\mu}^{ab}$ will map to 
smooth functions of the noncommutative coordinates $x^{i(0)}$ and $x^{{\tilde i}(0)}$. For the fields as the 
functions of non-commutative coordinates, we can introduce the Weyl ordering to define a suitable ordering for their 
products in the effective Lagrangian. This implies that the ordinary product should be replaced by the noncommutative 
$\star$-product. Here, the transition relation from the operator formalism for fields on noncommutative space to 
the representation in terms of ordinary function with the $star$-product reads as 
$[f,g]\rightarrow i\frac{\theta }{W}\epsilon^{rs}\partial_r f\partial_s g$, and 
$Tr(f_1\cdots f_n)\rightarrow \frac{W}{2\pi \theta }\int (f_1\star \cdots \star f_n)$. 
Finishing all of these, we find the effective action describing the fluctuations
\begin{equation}
I_{e}=\int d^3 \xi A_0 J^0 +\frac{\lambda \theta }{4\pi }\int d^3\xi\epsilon^{rst}
(A_r\star \partial_sA_t+\frac{2}{3}A_r\star A_s\star A_t),
\end{equation}
where $\lambda =4{\tilde \lambda}W$ and $J^0=\lambda /2$. 

The first term in the above equation is the chemical potential. The second term is the standard action
of the $U(1)$ noncommutative Chern-Simons theroy. Susskind \cite{Susskind} proposed this theory as the description of
the quantum Hall fluids on the plane. This Chern-Simons theory on the plane necessarily describes an infinite quantum
Hall system since the space noncommuattivity condition requires an infinite diemnsional Hilbert space. In other words,
the fields in this theory are infinite matrices corresponding to infinite number of electrons on the infinite plane. 
However, it is well known that Haldane's description of quantum Hall effect on the spherical geometry is equivalent 
to that of Laughlin's on the plane in the thermodynamic limit, taking the number of electrons large. Our conclusion is 
that in large $N$, the quantum Hall system on two-sphere is also described by the $U(1)$ noncommutative Chern-Simons 
theory. Physically, such a conclusion is reasonable.

It is, however, of interest to also describle finite systems of limited spatial extent with a finite number of electrons.
If we want to describe the quantum Hall fluids on 
two-sphere, we must regularize the noncommutative Chern-Simons theory for an infinite number of electrons. By means of
the Hopf fibration, the spinor $z=\left ( \begin{array} {l}z_1 \\
z_2 \end{array} \right )$ can be used to describe the dynamics of electrons on $S^2$. So, unlike the Polychronakos' \cite{
Polychronakos} finite matrix model on the plane, the regularized matrix model for particles on $S^2$ should correspond to
the $U(1)$ noncommutative Chern-Simons theory, and be described by the spinor matrix fields. It should be pointed out that
in such a model, the spinor $z$ must be regarded as a field with single particle rather than that with particles
$z_1$ and $z_2$.

\section{Regularized version of the $U(1)$ noncommutative Chern-Simons theory on $S^2$}

\indent

Now we describe the regularized version of the $U(1)$ noncommuattive Chern-Simons theory on $S^2$. This 
regularized matrix model should recover the $U(1)$ noncommutative Chern-Simons model in the large $N$ limit. Explicitly,
in the large $N$ limit, the equation of motion for $A_0$ as the constraint in the $U(1)$ noncommutative Chern-Simons 
model will provide the noncommutaivity of the coordinates, which equivalently produces the classical matrix commutator (9).
Such a regularized matrix model associated with the spinor matrix field $z$ can be obtained by following the 
Polychronakos' construction of the finite matrix model on the plane\cite{Polychronakos}. Notice that the Hopf mapping makes
the normal vector on $S^2$ be related to the coordinates of $S^3$ described by the spinor of two components
$z=\left ( \begin{array} {l}z_1 \\
z_2 \end{array} \right )$, i.e., $n^a =z^{\dagger}\sigma^a z$. This mapping relation is invariant under the $U(1)$ gauge
transformation $z\rightarrow z^{\prime}=e^{i\alpha}\left ( \begin{array} {l}z_1 \\
z_2 \end{array} \right )$. So it is natural for us to propose the following action
\begin{equation}
S_p =\frac{\lambda}{4}\int dt Tr\{iZ^{\dagger}D_t Z +2\theta A_0 -\omega Z^{\dagger}Z\}
+\frac{1}{2}\Psi^{\dagger}(i{\dot \Psi }-A_0 \Psi ) +h.c.
\end{equation}
to desccribe the finite number of electrons living on the two-dimensional sphere, where the covariant derivative is defined 
as $D_t=\partial_t +i[A_0,~~~]$. In the above equation, $\Psi $ and $Z$ are 
spinors with two components, defined by $\Psi =\left ( \begin{array} {l}\Psi_1 \\
\Psi_2 \end{array} \right )$ and $Z=\left ( \begin{array} {l}Z_1 \\
Z_2 \end{array} \right )$, respectively. $Z_\alpha ,\alpha =1,2$ are $N\times N$ complex matrices, $A_0$ is a $N\times N$
hermitian matrix, and $\Psi_{\alpha },\alpha =1,2$ are complex $N$-vectors. They, in the fundamental representation of 
the gauge group $U(N)$, are transformed as  
\begin{equation}
Z_{\alpha}\rightarrow UZ_{\alpha}U^{-1},~~~\Psi_{\alpha}\rightarrow U\Psi_{\alpha},~~~
A_0\rightarrow UA_0 U^{-1}+i{\dot U}U^{-1}.
\end{equation}
It is obvious that the action $S_p$ is invariant under the $U(N)$ gauge transformation.
Due to the gauge invariance of the action, we can choose gauge $A_0$ and impose the equation of motion of $A_0$ as the 
constraint
\begin{equation}
\frac{\lambda }{2}[Z,Z^{\dagger}]+\Psi \Psi^{\dagger}=\lambda \theta 
\end{equation}
which is from the variation of the action with respect to $A_0$. If we rescale $Z$ into $\sqrt{\frac{2}{\lambda }}
Z^{\prime}$ and denote $Z^{\prime}$ as $Z$, the above equation can be rewritten as
\begin{equation}
[Z,Z^{\dagger}]+\Psi \Psi^{\dagger}=\lambda \theta . 
\end{equation}

One can see from the action (28) that the conjugate momenta of $Z$ and $\Psi$ are $Z^{\dagger}$ and $\Psi^{\dagger}$, 
respectively. So they obey the classical matrix commutators $[(Z_{\alpha })_{mn},(Z_{\beta}^{\dagger})_{kl}]=-i\delta_{mk}
\delta_{nl}\delta_{\alpha \beta }$ and $[(\Psi_{\alpha })_{m},(\Psi _{\beta}^{\dagger})_{n}]=-i\delta_{mn}
\delta_{\alpha\beta }$. Since the spinor describes a particle moving on the two-dimensional sphere, we should regard
such spinor as a single oscillator. So there exist $N^2 +N$ uncoupled oscillators in the present system. Their 
hamiltonian is 
\begin{equation}
H=\omega Tr Z^{\dagger}Z=\omega \sum_{m,n,\alpha }(Z_{\alpha }^{\dagger})_{mn}(Z_{\alpha })_{nm}.
\end{equation}

The constraint equation can be used to reduce the space of quantum physical states. Since the constrained matrix
$G\equiv [Z,Z^{\dagger}]+\Psi \Psi^{\dagger}$ is the generator of unitary transformations of both $Z$ and $\Psi$, it 
must obey the commutation relations of the $U(N)$ algebra. In terms of the basis $\{I,T^a\}$ of the $U(N)$ algebra,
where $T^a$ are the $N^2-1$ normalized $SU(N)$ generators, the matrix fields $Z$ 
and $Z^{\dagger}$ can be expanded as
\begin{equation}
Z=z_0 +\sum_{a=1}^{N^2-1}z_a T^a,~~~~Z^{\dagger}=z_0^{\dagger} +\sum_{a=1}^{N^2-1}z_a^{\dagger} T^a .
\end{equation}
By using these expansions, we can express the constrained matrix as
\begin{equation}
G^a= Tr(GT^a)=-if^{abc}z_b^{\dagger}z_c +\Psi^{\dagger}T^a \Psi ,
\end{equation} 
where $f^{abc}$ are the structure constants of $SU(N)$ algebra, i.e., $[T^a ,T^b]=if^{abc}T^c$.

After quantization, the elements of the matrix fields $Z$, $Z^{\dagger}$ and the vector fields $\Psi$, $\Psi^{\dagger}$ 
become operators, and satisfy the fundamental commutation relations of operators 
\begin{equation}
[(Z_{\alpha })_{mn},(Z_{\beta}^{\dagger})_{kl}]=\delta_{mk}
\delta_{nl}\delta_{\alpha \beta },~~~~ 
[(\Psi_{\alpha })_{m},(\Psi _{\beta}^{\dagger})_{n}]=\delta_{mn}
\delta_{\alpha\beta }.
\end{equation}
Furthermore, after quantization, the expanded modes in the constrained matrix $G$ also become operators.
The constrained operators ${\hat G}^a$ satisfy the $SU(N)$ algebra, and can be regarded as the generators of the $SU(N)$ 
algebra. Because the generators $T^a$ of the $SU(N)$ algebra are traceless, the constrained matrix $G$ gives the traceless part
of the constraint equation (31). After quantization, the operators ${\hat G}^a$ become the projected operators of the 
quantum physical states in the matrix model
\begin{equation}
{\hat G}^a |Phys>=({\hat G}_Z^a +{\hat G}_\Psi^a)|Phys>=0 .
\end{equation}
On the other hand, the trace part of the constraint equation (31) produces the following constrained condition of the 
quantum physical states
\begin{equation}
(\Psi_n^{\dagger}\Psi_n -2N\lambda \theta )|Phys>=0.
\end{equation}
Since we are considering the matrix model of finite number of particles moving on the two-dimensional sphere, we must also
add the geometrical constraint to the quantum physical states to map the manifold parameterized by the coordinates $Z$ to
the two-sphere $S^2$. As mentioned above, $z^{\dagger}z=1$ together with the $U(1)$ gauge transformations of $z$, i.e.,
$z\rightarrow e^{i\alpha }z$, implies that the geometrical condition  $n^a n^a =1$ of $S^2$ is satisfied. However, in our
matrix model, this condition becomes
\begin{equation}
[Tr(Z^{\dagger}Z)-g]|Phys>_{s} =0 ,
\end{equation}
where $g$ is a parameter dependent of the model. Here $|Phys>_{s}$ stands for the geometrically stable configuration 
among the quantum physical states. In fact, the quantum physical states including the excitations do not belong to such
stable configurations, but the Laughlin-type states of quantum Hall fluids do.

From the constraint condition (36), we know that the physical states must be the singlet representations of the $SU(N)$ group, 
of which ${\hat G}^a$ are the generators. However, ${\hat G}_Z^a$ are only realized by the representations arising from  
products of the adjoint representations of $SU(N)$. Furthermore, $Z_1$ and $Z_2$ form  a spinor, and describe the spin
degree of freedom of particles. So they should appear in pairs in the singlet representations. Therefore, 
the representation of ${\hat G}_Z^a $ contains only the irreducible representations whose total number of boxes in their
Young tableau is an integer multiple of $2N$. Since the physical states are invariant under the sum of $G_Z^a$ and 
$G_{\Psi}^a$, the representations of $G_Z$ and $G_{\Psi}$ must be conjugate to each other so that their product contains the
singlet of the $SU(N)$ group. Hence, the irreduicble representations of $G_{\Psi}$ must also have a number of boxes which is
a multiple of $2N$. Following the Polychronakos' arguments\cite{Polychronakos}, from the other constraint condition (37), 
one knows that the number of boxes equals to the total number of operators of the spinor oscillators $\Psi^{\dagger}\Psi$. 
Thus, we conclude
\begin{equation}
\lambda \theta =k,
\end{equation}
where $k$ is an integer. This conclusion is the same as that of Polychronakos for the finite matrix model on the plane.
In fact, in the large $N$ limit, Haldane's quantum Hall effect model on the spherical geometry is equivalent to Laughlin's 
quantum Hall system on the plane. So, the level of $U(1)$ noncommutative Chern-Simons action is not changed by the geometry
on which the particles move.

In the Haldane's description of quantum Hall effect in terms of the spherical geometry, the spinors are the fundamental 
elements in the description of electrons on the two-dimensional sphere of which a Dirac monopole lies at the center, and are
the dynamical degrees of freedom of the electrons. On the other hand, it can be easily seen from the constraint conditions
(36) and (37) that only if the quantum physical states are the spin singlets the constraint of the $SU(N)$ invariance 
(36) is consistent with the vanishing condition of the total $U(1)$ charge (37).

To summarize, the quantum physical states of our matrix model must possess the following properties. (a) They are the 
singlet representaions of the $SU(N)$ group. (b) They must be the spin singlets. This implies that the same number of  
spin-up and spin-down components will be present in the quantum physical states. That is, they are the $SU(2)$ 
invariant states associated with the spin. (c) There exist $kN$ number of $\Psi^{\dagger}_1$ and $kN$ number of
$\Psi^{\dagger}_2$ in the 
quantum physical states, where $\Psi_1^{\dagger}$ and $\Psi_2^{\dagger}$ form the spinor $\Psi^{\dagger}$. (d) The 
geometrically stable states among the quantum physical states should also satisfy the geometrical constraint condition (38).

Subsequently, we shall determine the quantum physical states of the matrix model, which build up the physical Fock
space of the matrix model. Recall that the hamiltonian (32) of the system can be expressed as $\omega {\hat N}_Z$ in terms
of the number operator ${\hat N}_Z \equiv \sum_{m,n} Z^{\dagger}_{mn}Z_{nm}$ of the spinor oscillators $Z$. From this 
expression, we know that energy eigenstates will be linear combination of terms with a fixed number of $Z^{\dagger}$ 
creation operators acting on the Fock space vaccum $|0>$, which is defined by
\begin{equation}
Z_{mn}|0 \rangle =\Psi_{n}|0 \rangle =0 .
\end{equation}
The constraint conditions of quantum physical states require that all physical states must have a fixed number $Nk$  of
$\Psi_1^{\dagger}$ creation operators and the same number of $\Psi_2^{\dagger}$ creation operators acting on the Fock
vaccum. Furthermore, the number of $Z_1^{\dagger}$ creation operators appearing in the physical state should be the same as 
that of $Z_2^{\dagger}$. Thus, any physical  state describing an energy eigenstate will be a sum of terms of the form
\begin{equation}
\prod_{m=1}^{M}(Z_1^{\dagger})^{i_{m}}_{j_{m}}(Z_2^{\dagger})^{i_{m}^{\prime}}_{j_{m}^{\prime}}
\prod_{n=1}^{Nk}(\Psi_1^{\dagger})_{l_{n}}(\Psi_2^{\dagger})_{l_{n}^{\prime}} 
 |0 \rangle , 
\end{equation}
where, the fundamental indices of $SU(N)$ are written as upper indices, and the anti-fundamental indices as lower indices.

Now, the problem is how we can construct a singlet of both $SU(N)$ and spin from the equation (41) through 
contracting all indices by the covariant tensors of $SU(N)$. Since the product of an upper indix epsilon tensor
and a power index epsilon tensor may be rewritten as a sum of products of the delta functions, we may only use one type of 
epsilon tensor to finish the contraction. To continue the construction of the quantum physical states, first of all,
we shall establish a few of lemmas, which are the generalizations of the facts given by Hellerman and Raamsolonk in the
appendix of their paper\cite{Hellerman}.

\noindent {\bf Lemma 1}: Seting $\chi (u,v)\equiv \prod_{i<j} (u_{i}v_{j} -u_{j}v_{i})$ where $u$ and $v$ are two components of 
spinor $z$, i.e., $z=\left ( \begin{array} {l}u \\
v \end{array} \right )$, we have 
\begin{equation}
\chi (u,v)=\epsilon^{i_{1}\cdots i_{N}}\prod_{n=1}^{N}(u^{N-n}v^{n-1})_{i_{n}},
\end{equation}
where we have abbreviated $\epsilon^{(i{\tilde i})_{1}\cdots i{\tilde i}_{N}}\prod_{n=1}^{N}(u^{N-n})_{i_{n}}
(v^{n-1})_{{\tilde i}_{n}}$ as $\epsilon^{i_{1}\cdots i_{N}}\prod_{n=1}^{N}(u^{N-n}v^{n-1})_{i_{n}}$.

\noindent Proof: From the definition of $\chi (u,v)$ and its expression (42), one can see that they all are completely antisymmetric,
and have the same order of $u$ and $v$ power $N(N-1)$. hence, the definition of $\chi (u,v)$  must be equals its exprssion
up to a numerical factor. Taking a fixed $N$, e.g., $N=3$, we can check thst the numerical factro is equal to 1.

\noindent {\bf Lemma 2}: Any polynomial $D(u,v)=\epsilon^{i_{1}\cdots i_{N}}\prod_{m=1}^{N}(u^{n_m}v^{{\tilde n}_m})_{i_{m}}$, 
where $\sum_{i=1}^{N}n_{i}=\sum_{i=1}^{N}{\tilde n}_{i}$, may be written as a sum of terms of the form
\begin{equation}
F(u,v)=\prod_{n=1}^{N}S_{n}^{c_{n}}{\tilde S}_{n}^{{\tilde c}_{n}}\chi (u,v),
\end{equation}
here $S_l =\sum_{i=1}^{N}u_i^l$ and ${\tilde S}_l =\sum_{i=1}^{N}v_i^l$. The equality $\sum_{i=1}^{N}n_{i}=
\sum_{i=1}^{N}{\tilde n}_{i}$ is the conclusion of spin singlet, which ipmlies that $\sum_{i=1}^{N}ic_{i}=
\sum_{n=1}^{N}i{\tilde c}_{i}$.

\noindent {\bf Lemma 3}: Let $\Psi_1^{\dagger}$, $\Psi_2^{\dagger}$ and $Z_1^{\dagger}$, $Z_2^{\dagger}$ be the $N$-dimensional vectors 
and the $N\times N$ matrices of commuting variables, respectively. Thus, any expression of the form
\begin{eqnarray}
{\cal D}(\Psi_1^{\dagger},\Psi_2^{\dagger};Z_1^{\dagger},Z_2^{\dagger})& = &
\epsilon^{(i{\tilde i})_{1}\cdots (i{\tilde i})_{N}}\prod_{l=1}^{N}(\Psi_1^{\dagger}Z_1^{\dagger n_l}
Z_2^{\dagger {\tilde n}_l}\Psi_2^{\dagger})_{(i{\tilde i})_l} \nonumber\\
& \equiv & \epsilon^{i_{1}\cdots i_{N}}\prod_{l=1}^{N}(\Psi_1^{\dagger}Z_1^{\dagger n_l}
Z_2^{\dagger {\tilde n}_l}\Psi_2^{\dagger})_{i_l} 
\end{eqnarray}
may be uniquely written as a sum of terms of form
\begin{equation}
{\cal F}(\Psi_1^{\dagger},\Psi_2^{\dagger};Z_1^{\dagger},Z_2^{\dagger})=  
\prod_{i=1}^{N}(TrZ_1^{\dagger i})^{c_i}(TrZ_2^{\dagger i})^{{\tilde c}_i}
\epsilon^{i_{1}\cdots i_{N}}\prod_{n=1}^{N}(\Psi_1^{\dagger}Z_1^{\dagger N-n}
Z_2^{\dagger n-1}\Psi_2^{\dagger})_{i_n} .
\end{equation}
conversely, the conclusion of the lemma also is true.

The proofs of the lemma 2 and the lemma 3 are completely parallel to those of the fact 2 and the fact 3 provided by
Hellerman and Raamsolonk. Here we shall omit them. The above lemmas are the base of constructing the quantum physical
states of the matrix model presented here.

Now, we return to the contraction of the Fock states (41). The problem is to determine all ways of combining the 
symmetrized anti-fundamentals $\Psi_1^{\dagger}$, $\Psi_2^{\dagger}$ with any fixed number of adjoints to form a singlet of
$SU(N)$. Unlike the Polychronakos' finite matrix model, this singlet of $SU(N)$ is also a spin singlet. Let us notice
that $Z_1^{\dagger}, \Psi_1^{\dagger}$ should be regarded as the spin-up components and $Z_2^{\dagger}, 
\Psi_2^{\dagger}$  as spin-down components. In order to avoid the spin confusion, we shall make $Nk$ creation 
operators $\Psi_1^{\dagger}$ contract with a fixed number of creation operators $Z_1^{\dagger}$ and $Nk$ operators
$\Psi_2^{\dagger}$ with certain number of $Z_2^{\dagger}$ to form a singlet of $SU(N)$. Precisely, let us consider first
the indices 
of the $Nk$ number of $\Psi_1^{\dagger}$. The lowr index on each $\Psi_1^{\dagger}$ must contract either with the upper index on a
$Z_1^{\dagger}$ or with an epsilon tensor. If the $\Psi_1^{\dagger}$ contracts with a $Z_1^{\dagger}$, the resulting
object will again have a single lower index, i.e., $(\Psi_1^{\dagger})_{i_1}(Z_1^{\dagger})^{i_1}_{j_1}\rightarrow 
(\Psi_1^{\dagger}Z_1^{\dagger})_{j_1}$. The lower index on the object may again contract either with the upper index on a 
$Z_1^{\dagger}$ or with an epsilon tensor. Repeating this logic, we conclude that each $\Psi_1^{\dagger}$ will contract 
with some number of $Z_1^{\dagger}$ and that the resulting object will have its single lower index contracted with an
upper index epsilon tensor. So the result is $(\Psi_1^{\dagger}Z_1^{\dagger n_1})_{i_1}$. Similarly, we have
$(\Psi_2^{\dagger}Z_2^{\dagger {\tilde n}_1})_{{\tilde i}_1}$. However, the indices of 
$(\Psi_1^{\dagger}Z_1^{\dagger n_1})_{i_1}$ and $(\Psi_2^{\dagger}Z_2^{\dagger {\tilde n}_1})_{{\tilde i}_1}$ 
belong to the same particle lable since $Z_1^{\dagger}$ and $Z_2^{\dagger}$ are associated with the componetnts of spin up 
and spin down, respectively, of the particle. This implies that for the fixed particle, $\Psi_1^{\dagger}Z_1^{\dagger n_1}$ 
and $\Psi_2^{\dagger}Z_2^{\dagger {\tilde n}_1}$ should appear in one contracted element,
i.e., $(\Psi_1^{\dagger}Z_1^{\dagger n_1}Z_2^{\dagger{\tilde n}_1}\Psi_2^{\dagger })_{(i{\tilde i})_1}$. Such $N$ elements 
are contracted with the upper indices of an epsilon tensor to produce the fundamental contraction block
\begin{equation}
\epsilon^{i_1\cdots i_N}\prod_{l=1}^{N} 
(\Psi_1^{\dagger}Z_1^{\dagger n_l}Z_2^{\dagger {\tilde n}_l}\Psi_2^{\dagger})_{i_l},
\end{equation}
where we have admitted the abbreviated symbol appeared in the lemmas.

Because there exist $Nk$ number of $\Psi_1^{\dagger}$ and $Nk$ number of $\Psi_2^{\dagger}$ in the quantum physical states, the physical
states are composed of $k$ fundamental contraction blocks. So, using the lemmas mentioned by us above, we can write the 
minimal basis of the physical energy eigenstates being both the $SU(N)$ singlets and the spin singlets as
\begin{equation}
|\{n_i\},\{{\tilde n}_i\},k \rangle =
\epsilon^{i_1\cdots i_N}\prod_{l=1}^{N} 
(\Psi_1^{\dagger}Z_1^{\dagger n_l}Z_2^{\dagger {\tilde n}_l}\Psi_2^{\dagger})_{i_l} 
(\epsilon^{i_{1}\cdots i_{N}}\prod_{n=1}^{N}(\Psi_1^{\dagger}Z_1^{\dagger N-n}
Z_2^{\dagger n-l}\Psi_2^{\dagger})_{i_n})^{k-1} |0 \rangle ,
\end{equation}
where $\{n_i\}$ and $\{{\tilde n}_i\}$ satisfy the relation $\sum_i n_i =\sum_i {\tilde n}_i$ from the requirement of the
spin singlet. On the other hand, there exist some additional $Z_i^{\dagger}$, for $i=1,2$, which are contracted amongst 
themselve as the terms of products $Tr(Z_1^{\dagger}), \cdots Tr(Z_1^{\dagger N})$ and 
$Tr(Z_1^{\dagger}), \cdots Tr(Z_1^{\dagger N})$ to form the physical states. By means of the Lemma 3, they can be used to 
build up another set of minimal basis of the physical energy eigenstates as 
\begin{equation}
|\{c_i\},\{{\tilde c}_i\},k \rangle =
\prod_{i=1}^{N}(TrZ_1^{\dagger i})^{c_i}(TrZ_2^{\dagger i})^{{\tilde c}_i}
(\epsilon^{i_{1}\cdots i_{N}}\prod_{n=1}^{N}(\Psi_1^{\dagger}Z_1^{\dagger N-n}
Z_2^{\dagger n-1}\Psi_2^{\dagger})_{i_n})^k  |0 \rangle .
\end{equation}
The spin singlet condition of the physical states leads to $\sum_{i=1}^N ic_i =\sum_{i=1}^N i{\tilde c}_i$. By using 
the expression of hamiltonian (32), we can easily read off the energy eigenvalues of the above bases. The states of the
former basis have the structure of energy levels as the following
\begin{equation}
E(\{n_i\},\{{\tilde n}_i\},k)=\omega [(k-1)N(N-1) +\sum_i n_i +\sum_i {\tilde n}_i ]
=\omega [(k-1)N(N-1) +2\sum_i n_i  ] .
\end{equation}
The energy eigenvalues of the latter basis are given by
\begin{equation}
E(\{c_i\},\{{\tilde c}_i\},k )=\omega [kN(N-1) +\sum_{i=1}^N ic_i +\sum_{i=1}^N i{\tilde c}_i] 
=\omega [kN(N-1) +2\sum_{i=1}^N ic_i ] .
\end{equation}

Furthermore, from the expressions (47) and (48) of the minimal bases of the physical energy eigenstates, we find that the 
physical ground state of the present finite matrix model is expressed by
\begin{equation}
|\{0\},\{0\},k \rangle =
(\epsilon^{i_{1}\cdots i_{N}}\prod_{n=1}^{N}(\Psi_1^{\dagger}Z_1^{\dagger N-n}
Z_2^{\dagger n-1}\Psi_2^{\dagger})_{i_n})^k  |0 \rangle
\equiv L^{\dagger k} |0 \rangle .
\end{equation}

Roughly speaking, after finishing the formal substitutions $\Psi_1^{\dagger}\rightarrow 1$, $\Psi_2^{\dagger}\rightarrow 1$ and 
$Z^{\dagger}_{1ij}\rightarrow \delta_{ij}A_{1j}$, $Z^{\dagger}_{1ij}\rightarrow \delta_{ij}A_{1j}$ in (51), we can get 
$|0,k \rangle =(\epsilon^{i_{1}\cdots i_{N}}\prod_{n=1}^{N}(A^{1\dagger N-n}A^{2\dagger n-1})_{i_{n}})^{k}|0 \rangle $. 
Furthermore, if $u$ and $v$ are regarded as the eigenvalue parameters of $A^{1}$ and $A^{2}$ in the coherent state picture 
respectively, we find that $\langle u,v |0,k\rangle =(\epsilon^{i_{1} \cdots i_{N}}\prod_{n=1}^{N}(u^{N-n}v^{n-1})_{i_{n}} )^{k} 
=\prod_{i<j} (u_{i}v_{j} -u_{j}v_{i})^{k}$. It is just the same as the ground state wavefunction of two-dimensional
quantum Hall fluid on the Haldane's spherical geometry \cite{Haldane}. However, as pointed by Polychronakos \cite{Polychronakos}, 
the classical value of the inverse filling fraction is shifted quantum mechanically if one uses the finite matrix 
Chern-Simons theory to describe the fractional quantum Hall states. This can be equivalently viewed as a renormalization of 
the Chern-Simons coefficient. In fact, this level shift of the matrix Chern-Simons model can be read off from the well 
known quantum mechanical level shift of the corresponding Chern-Simon theory. The renormalization of the level of 
Chern-Simons theory has been finished by using a biparameter family of BRS invariant regularization methods of Chern-Simons 
theory \cite{Giavarini}. This renormalization leads to the level shift $k\rightarrow k+sign(k)c_{V}$, where $k$ is 
the bare Chern-Simons level parameter and $c_{V}$ the quadratic Casimir in the adjoint representation of the gauge 
group of Chern-Simons theory. As mentioned previously, our finite Chern-Simons matrix model on $S^2$ corresponds to the
$U(1)$ Chern-Simons theory. Hence, physical states in the matrix model presented here at level $k$ should be identified with the quantum Hall 
states at the filling fraction $1/(k+1)$ rather than $1/k$, like the Polychronakos' finite matrix model on the plane. 
That is, the Laughlin type wavefunction on the two-dimensional Haldane's spherical geometry for 
filling fraction $1/(k+1)$ can be equivalently described by the physical ground state of the finite Chern-Simons matrix 
model on $S^2$.

The reason that the state (51) is regarded as the physical ground state becomes clear from the following discussion. The 
system considered by us is a priori $2N(N+1)$ uncoupled oscillators, which are composed of $2N^2$ harmonic oscillators
coordinated by $Z_1$, $Z_2$ and $2N$ harmonic oscillators done by $\Psi_1$, $\Psi_2$. However, they should be regarded as
$N(N+1)$ uncoupled spinor osillators since $Z=\left ( \begin{array} {l}Z_1 \\
Z_2 \end{array} \right )$ and $\Psi =\left ( \begin{array} {l}\Psi_1 \\
\Psi_2 \end{array} \right )$ must be viewed as the spinors describing the particles on $S^2$. Furthermore, what couples the 
spinor oscillators is $N^2-1$ constraint equations in the traceless part of the Gauss constraint (31). Effectively, we can
describe the system with $N+1$ independent oscillators. All $SU(N)$ invariant states can be spanned by the operators,
$Q_{1n}^{\dagger}=Tr(Z_1^{\dagger n})$, $Q_{2n}^{\dagger}=Tr(Z_2^{\dagger n})$ with $n=1,2,\cdots,N$, and $L^{\dagger k}$
acting on the Fock vaccum. However, the spin singlet condition of physical states must result in the balance of the numbers
of operators $\{Q_{1n}^{\dagger}\}$ and $\{Q_{2n}^{\dagger}\}$ appearing in the physical states. So they can be regarded as 
the $N$ independent spinor osillators together with the operator $L^{\dagger k}$ composed of the $N+1$ independent 
oscillators physically describing the system. A useful conclusion in mathematics is that the operators 
$Q_{1l}^{\dagger}$ and $Q_{2l}^{\dagger}$ for $l>N$ can be expressed as the homogeneous polynomials of total order $l$ in      
$\{Q_{11}^{\dagger}, Q_{12}^{\dagger},\cdots,Q_{1N}^{\dagger}\}$ and $\{Q_{21}^{\dagger}, Q_{22}^{\dagger},\cdots,
Q_{2N}^{\dagger}\}$, respectively, with constant coefficients which are common to all operators. Based on this conclusion and 
the commutation relations between $Z$ and $Z^{\dagger}$, we have $Q_{1l}Q_{2l}L^{\dagger k}|0 \rangle =0,~~for~~all~~l$. 
This means that the state $L^{\dagger k}|0 \rangle \equiv |0,k \rangle $ is the physical vaccum with respect to all 
operators $Q_{1l}$ and $Q_{2l}$. Equivalently, the Laughlin-type state $|0,k \rangle $ is the physical ground state of our
finite matrix model. In the next section, we shall discuss the excitation states produced by the creation operators
$Q_{1l_1}^{\dagger}$ and $Q_{2l_2}^{\dagger}$ acting on the ground state.

\section{Quasiparticle excitations and hierarchy of the quantum Hall fluids on $S^2$ in the finite matrix model}

\indent

The low-lying excitations in our matrix model can be described in terms of quasiparticles and quasiholes by following
\cite{Polychronakos,Jonke}. A quasiparticle state is obtained by peeling a 'particle' from the surface of the Fermi sea. 
That is, one quasiparticle obtained by exciting a 'particle' at Fermi level by energy amount $n\omega$ is described by
\begin{eqnarray}
p^{1\dagger}_{n}|0,k \rangle =&(\epsilon^{i_{1}\cdots i_{N}}& \prod_{m=1}^{N} 
(\Psi_1^{\dagger}Z_1^{\dagger N-m}
Z_2^{\dagger m-1}\Psi_2^{\dagger})_{i_{m}}
)^{k-1} \nonumber\\
&\epsilon^{i_{1}\cdots i_{N}} &
(\Psi_1^{\dagger}Z_1^{\dagger N-1+n}Z_2^{\dagger 0}\Psi_2^{\dagger})_{i_{1}} 
\prod_{m=2}^{N}(\Psi_1^{\dagger}Z_1^{\dagger N-m}Z_2^{\dagger m-1}\Psi_2^{\dagger})_{i_{m}}|0\rangle .
\end{eqnarray}
The quasiholes correspond to the minimal excitations of the ground state inside the quantum Hall fluid. One quasihole 
excitation is obtained by creating a gap inside the QHF with the energy increase $m\omega$
\begin{eqnarray}
h^{1\dagger}_{m}|0, k \rangle =&(\epsilon^{i_{1}\cdots i_{N}}& \prod_{n=1}^{N} 
(\Psi_1^{\dagger}Z_1^{\dagger N-n}
Z_2^{\dagger n-1}\Psi_2^{\dagger})_{i_{n}}
)^{k-1} \nonumber\\
&\epsilon^{i_{1}\cdots i_{N}}&\prod_{n=1}^{m}
(\Psi_1^{\dagger}Z_1^{\dagger N-n+1}Z_2^{\dagger n-1}\Psi_2^{\dagger})_{i_{n}} 
\prod_{n=m+1}^{N}(\Psi_1^{\dagger}Z_1^{\dagger N-n}Z_2^{\dagger n-1}\Psi_2^{\dagger})_{i_{n}}|0\rangle .
\end{eqnarray}
Obviously, $p^{1\dagger}_{1}=h^{1\dagger}_{1}$. So there is no fundamental distinction between 'particles' and 'holes' in 
the matrix model. Similarly, one can describe the quasiparticle $p^{2\dagger}_{n}$ and quasihole $h^{2\dagger}_{m}$ of 
excitations corresponding to the oscillator field $Z_{2}$. They read as
\begin{eqnarray}
p^{2\dagger}_{n}|0,k \rangle =&(\epsilon^{i_{1}\cdots i_{N}}&\prod_{m=1}^{N} 
(\Psi_1^{\dagger}Z_1^{\dagger N-m}
Z_2^{\dagger m-1}\Psi_2^{\dagger})_{i_{m}}
)^{k-1} \nonumber\\
&\epsilon^{i_{1}\cdots i_{N}} &
\prod_{m=1}^{N-1}(\Psi_1^{\dagger}Z_1^{\dagger N-m}Z_2^{\dagger m-1}\Psi_2^{\dagger})_{i_{m}}
(\Psi_1^{\dagger}Z_1^{\dagger 0}Z_2^{\dagger N-1+n}\Psi_2^{\dagger})_{i_{1}} |0\rangle ,
\end{eqnarray}
and
\begin{eqnarray}
h^{2\dagger}_{m}|0, k \rangle =&(\epsilon^{i_{1}\cdots i_{N}}& \prod_{n=1}^{N}
(\Psi_1^{\dagger}Z_1^{\dagger N-n}
Z_2^{\dagger n-1}\Psi_2^{\dagger})_{i_{n}}
)^{k-1} \nonumber\\
&\epsilon^{i_{1}\cdots i_{N}}& 
\prod_{n=1}^{N-m}(\Psi_1^{\dagger}Z_1^{\dagger N-n}Z_2^{\dagger n-1}\Psi_2^{\dagger})_{i_{n}}
\prod_{n=N-m+1}^{N}
(\Psi_1^{\dagger}Z_1^{\dagger N-n}Z_2^{\dagger n}\Psi_2^{\dagger})_{i_{n}} |0\rangle .
\end{eqnarray}

Although all of these excitations are the fundamental excitations in the finite matrix model here, they can not 
be regarded directly as the physical low-lying excitations in the matrix model. The physical exciting states must obey 
the constraint condition of physical states of the matrix model, which is just the spin singlet condition of the physical 
states. By using the lemmas shown by us in the previous section, one can find that all of the fundamental excitations 
as mentioned above can be equivalently expressed in terms of the following states
\begin{equation}
P_{n_1,n_2}^{\dagger}|0, k \rangle =
\prod_{j=1}^{N}(Tr Z_1^{\dagger j})^{c_{j}}(Tr Z_2^{\dagger j})^{{\tilde c}_j} 
|0, k \rangle =\prod_{j=1}^{N}(Q_{1j}^{\dagger})^{c_{j}}(Q_{2j}^{\dagger})^{{\tilde c}_j} 
|0, k \rangle ,
\end{equation}
where, $n_1=\sum_{i=1}^{N}ic_{i}$ and $n_2=\sum_{i=1}^{N}i{\tilde c}_{i}$. Indeed, they are exciting states with respect 
to the Laughlin-type ground state $|0,k \rangle $ clearly from the discussion of the last paragraph in the above section. 
The physical excitation states can be constructed by using the expression (56) of fundamental excitations and by adding the
spin singlet condition of physical states to restrict them.

Following Haldane \cite{Haldane}, we can construct the collective ground state of two-dimensional quantum Hall fluid from the
present matrix model by the condensing of the quasiparticle and quasihole excitations. The condensation here means that the
physical state including the excitations becomes the Laughlin-type fluid state, like the physical ground state
$|0,k \rangle $. This Laughlin-type fluid state is given by
\begin{equation}
|0, p_1, k \rangle =
(\epsilon^{i_1,\cdots,i_{N_1}}
\prod_{n=1}^{N_1}(P^{1\dagger N_1 -n}P^{2\dagger n-1})_{i_n}))^{p_1}
|0, k \rangle ,
\end{equation}
where $N_1=N/p_1 +1$ and $N$ is divisible by $p_1$. $P^{1\dagger i}$ and $P^{2\dagger i}$ stand for $P_{i,0}^{\dagger}$ and
$P_{0,i}^{\dagger}$, respectively, defined by the expression (56). In fact, the condition $N_1=N/p_1 +1$ is required by 
the fact that
if one uses the operators $\{Q_{1n}^{\dagger}\}$ and $\{Q_{2n}^{\dagger}\}$ to be a set of independent creation operators, the 
condition of operator powers $n\leq N$ must be satisfied, otherwise, the constructed creation operators will be dependent. 
By using the lemmas in the section three, one can easily check that the excitation fluid state indeed satisfies the spin
singlet condition of the physical states in the matrix model.

We can also construct the excitation states of the two-dimensional excitation fluid in a way similar to the construction
of the excitation states of the quantum Hall fluid state $|0,k \rangle $. It will be convenient to use the latter basis of
the Fock space of the matrix model to do this. Introducing operators $P_s^{1\dagger j}=\sum_{n=1}^{N_1}P_n^{1\dagger j}$
and $P_s^{2\dagger j}=\sum_{n=1}^{N_1}P_n^{2\dagger j}$, we can write the excitation states of the excitation fluid as
\begin{equation}
|\{c_{1i}\},\{{\tilde c}_{1i}\}, p_1, k \rangle =
\prod_{j=1}^{N_1}(P_s^{1\dagger j})^{c_{1j}}(P_s^{2\dagger j})^{{\tilde c}_{1j}}|0,p_1 ,k \rangle .
\end{equation}
Although these states do not generally satisfy all of constraint conditions of the physical states in the matrix model, 
they are the mediate states to construct the physical ground state of the excitation fluid in the next level obeying the 
constraint conditions. Here, the further constraint condition of the physical excitation states is the spin singlet
condition. 
One can get the next level of the Laughlin-type fluid state with the excitations derived by the excitation fluid by 
further condensing the 'quasiparticle' and 'quasihole' excitations from the excitation fluid. Furthermore,
similar to Haldane's hierarchical scheme of the quantum Hall fluids, in our matrix model the procedure of constructing 
the two-dimensional excitation fluids can be iterated, and leads to the hierarchy of the two-dimensional quantum Hall 
fluid states. We give the result of the iterated construction of the quantum Hall states as following
\begin{equation}
|0, p_{m},\cdots,p_1, k \rangle =
\prod_{q=1}^{m}(\epsilon^{i_1,\cdots,i_{N_q}} 
\prod_{n=1}^{N_q}(P_{q-1}^{1\dagger N_q -n}P_{q-1}^{2\dagger n-1})_{i_{n}})^{p_q}|0, k \rangle ,
\end{equation}
where the iterated relation is given by $p_q (N_q-1)+N_{q+1}=N_{q-1}$ with $N_{q}=0$ for $q>m$ and $N_{0}=N$. The fundamental 
excitation operators of the excitation fluids $P_m^{1\dagger j}\equiv P_{j,0}^{(m)}$ and 
$P_m^{2\dagger j}\equiv P_{0,j}^{(m)}$ are determined by the set of equations
\begin{equation}
P_{n_1,n_2}^{(m)}|0, p_{m},\cdots,p_1, k \rangle =
\prod_{j=1}^{N_m}(P_{s,m-1}^{1\dagger j})^{c_j}(P_{s,m-1}^{2\dagger j})^{{\tilde c}_j}|0, p_{m},\cdots,p_1, k \rangle ,
\end{equation}
where $n_1=\sum_{j=1}^{N_m}jc_j $ and $n_2=\sum_{j=1}^{N_m}j{\tilde c}_j $. The symmetric operators in the equation (60)
are defined as $P_{s,m-1}^{1\dagger j}=\sum_{i=1}^{N_m}(P_{m-1}^{1\dagger j})_i$ and $P_{s,m-1}^{1\dagger j}=
\sum_{i=1}^{N_m}(P_{m-1}^{1\dagger j})_i$. In fact, in the procedure of constructing the hierarchy of two-dimensional quantum 
Hall fluids in the matrix model, we have finished the construction of all quasiparticle and quasihole excitations of the 
quantum Hall excitation fluids.

Haldane's idea for the hierarchy of fluid states is to consider a slightly different field strength with parent field 
strength. The hierarchical fluid states are the low-energy states at this field strength, which can be considered as derived 
from the fluid state described by the Laughlin-type state at the parent field strength with an imbalance of quasiparticle 
and quasihole excitations. In our matrix model, the physical states $|0, p_{m},\cdots,p_1, k \rangle $ correspond to Haldane's
low-energy states. Physically, the imbalance of quasiparticle and quasihole excitations leads to that the degeneracy of the
lowest Landau level states in the hierarchical fluids becomes
\begin{equation}
2S+1=k(N-1)+N_1 +1 ,
\end{equation}
where $N_1$ in the expression of degeneracy is from the contribution of the imbalance of quasiparticle and 
quasihole excitations with respect to the background of the quantum Hall fluid for the electrons. Such $N_1$ can be obtained 
by solving the hierarchical iterated equations $p_q(N_q-1)+N_{q+1}=N_{q-1}$. The result is
\begin{equation}
N_1= \frac{N}{p_1+\frac{1}{p_2+\cdots+\frac{1}{p_m}}}
+\frac{1}{p_1+\frac{1}{p_2+\cdots+\frac{1}{p_m}}}[p_1 -\frac{1}{p_2+\frac{1}{p_3+\cdots+\frac{1}{p_m}}}[
\cdots [p_{m-2}-\frac{p_{m-1}-1}{p_{m-1}+\frac{1}{p_m}}]]].
\end{equation}
The filling factor means the lowest Landau level occupation factor. For the general hierarchical fluid states, the imbalance 
of excitations with respect to the background of the electron's quantum Hall fluid can be affected by the imbalances of the
excitations of the excitation fluids. This results in different $N_1$ for the variation of hierarchical fluid states.
But the spin of electrons in the hierarical fluid state is given by the formula $\frac{1}{2}{\tilde k}(N-1)+\frac{1}{2}N_1$,
where we have involved in the level shifting $k\rightarrow {\tilde k}=k+1$ of the Chern-Simons matrix model, while the 
number of electrons is still $N$. So we obtain the filling factor of the m-th hierarchical fluid state in the thermodynamic 
limit
\begin{equation}
\nu=\lim_{N\rightarrow \infty }\frac{N}{{\tilde k}(N-1)+N_1 +1}=
\frac{1}{{\tilde k}+\frac{1}{p_1+\frac{1}{p_2+\cdots+\frac{1}{p_m}}}}.
\end{equation}

Since the hierarchical quantum Hall fluid states are directly from the condensation of excitations of the electron's 
quantum Hall fluid in the matrix model, we find the energies of these hierarchical fluid states by using the 
hamiltonian expression of the matrix model. Explicitly,
\begin{equation}
H|0, p_{m},\cdots,p_1, k \rangle =
\omega [kN(N-1)+\sum_{q=1}^{m}p_q N_q (N_q -1)]|0, p_{m},\cdots,p_1, k \rangle .
\end{equation}
Substituting the iterated equations $p_q(N_q-1)+N_{q+1}=N_{q-1}$, we obtain the energy of the $m$-th hierarchical
fluid state 
\begin{equation}
E(p_m ,\cdots,p_1 ,k)=\omega [kN(N-1)+NN_1].
\end{equation}
The term for $N_1$ in the energy is from the contribution of condensing of the excitations. This implies that these 
hierarchical fluid states are the substable states of the matrix model since their energies are higher than that of 
the parent fluid state, i.e., the physical ground state of the matrix model. It should be emphasized that the hierarchical 
quantum Hall fluids are dynamically formed by condensing the 'quasiparticle' and 'quasihole' excitations level by level. 
In other words, there consistently exist such hierarchical fluid states in our matrix model without any requirement of 
modifications of the matrix model.

\section{Summary and outlook}

\indent

If one considers particle's motion on two-sphere in a radial monopole magnetic field, the configurations of particle's 
coordinates can not be smoothly defined globally over the entire $S^2$ due to the singularity of the Dirac monopole.
The effective action of the system can be described in a singlularity-free way by using a nontrivial bundle over $S^2$, 
which can be obtained by the Hopf fibration with base $S^2$. The existence of the Dirac monopole in the 2-dimensional 
quantum Hall system makes the coordinates of particles moving on the two-sphere become noncommutative. The appearance of 
such monopole also results in the irreducible representations of $SU(2)$ belonging to the Hilbert space composed of the 
lowest Landau level states of the system to be truncated. This phenomenon occurs in the 
description of fuzzy two-sphere. We have explicitly shown that the noncommutative structure of fuzzy $S^2$ appears indeed
in the Haldane's model of quantum Hall effect by restricting to the lowest Landau level states. In order to establish the
description of noncommutative field theory for the quantum Hall syatem on $S^2$, we have provided the Hopf mapping of the 
fuzzy $S^2$, i.e. (14) and (15). 
This mapping between the fuzzy manifolds plays the essential role in the descrption of noncommutative
field theory of quantum Hall fluids on $S^2$. It results in that the finite matrix model of quantum Hall fluids on $S^2$ is 
related to the matrix fields of the spinor with two complex components, which are from the matrix regularized-version of 
the spinor in the original Hopf map. In the first Hopf map, the Hopf fibration of $S^2$ can be regarded as a principal fibre 
bundle with the base space $S^2$ and a $U(1)$ structure group. This $U(1)$ gauge group is iterated in the formulation of 
noncommutativ field theory of the quantum Hall fluids on $S^2$. This implies that the finite matrix model given by us is the
finite matrix regularized version of the $U(1)$ noncommutative Chern-Simons theory on $S^2$. The finite matrix model (28) on 
$S^2$ involves in the matrix and vector fields of the spinor, different from the Polychronakos'
finite matrix model on the plane. In fact, the Hopf mapping of the fuzzy manifolds related to the second Hopf map is very important for the 
descriptions of the quantum Hall effect on $S^4$ and of the open two-brane in M theory\cite{Zhang,Chen}. It should be 
interesting to further recognize the mathematical implication of the Hopf mapping between the fuzzy manifolds and their
applications in physics.
 
Hellerman and Raamsdonk \cite{Hellerman} had speculated the second-quantized field theoretical description of 
the quantum Hall fluid for various filling fractions, which is given by the regularized matrix version of the 
noncommutative $U(1)$ Chern-Simons theory on the plane, by determining the completely minimal basis of exact wavefunctions 
for the Polychronakos' finite matrix model. For the quantum Hall fluids on $S^2$, we have determined the complete set of 
the physical quantum states of the finite matrix model on $S^2$, and shown the correspondence between the physical
ground state of our model and the Laughlin-type wavefunction on the Haldane's spherical geometry. Although such 
a determination of physical states is the generalization of Hellerman and Raamsdonk's work for the finite matrix model
on the plane, it is nontrivial because the spinor matrix fields and the spinor vector fields are included in the finite 
matrix model on $S^2$. On the other hand, we have attempted to establish the second-quantized field theoretical description of 
the quantum Hall fluids for various filling fractions on the Haldane's spherical geometry, and determined some essential 
physical elements in the quantum Hall fluid. However, further investigations are needed. For example, Karabali and 
Sakita\cite{Karabali} used the technique of coherent state to realize the Laughlin wave functions on the plane in the
Polychronakos' finite matrix model on the plane. The technique of $SU(2)$ coherent states can be used to 
investigate the explicit relation between the physical states and the Haldane's quantum Hall wavefunctions.

It is an interesting conclusion of our work that the hierarchical Hall fluid states can be dynamically generated in the 
finite matrix model on $S^2$. The formation of these hierarchical Hall fluid states originates from the condensing of
excitations of the quantum Hall fluids level by level. This dynamical mechanism is consistent with the original idea of 
Haldane's hierarchy. In the procedure of the construction of the hierarchical Hall fluid states, we have found the explicit 
forms of 'quasiparticle' and 'quasihole' excitations in each level of the hierarchy. We believe that these results are 
helpful in studying the correlations of excitations in the quantum Hall effect and the interaction behaviour of them.  
\\

This work was partly supported by Austrialian Research Council. The work of Y.X.C. was also partly supported by the NNSF of 
China (Grant No.90203003), the Foundation of Education Ministry of China (Grant No.010335025) and the Tang Yong-Qian 
Foundation of Zhejiang University.


\end{document}